\documentclass[twocolumn,english,aps]{revtex4}
\usepackage[T1]{fontenc}
\usepackage[latin1]{inputenc}
\usepackage{babel}

\makeatletter

\providecommand{\LyX}{L\kern-.1667em\lower.25em\hbox{Y}\kern-.125emX\@}
\newcommand{\boldsymbol}[1]{\mbox{\boldmath $#1$}}

\makeatother
\begin{document}

\title{Dynamical instability of a condensate induced by a rotating thermal
gas}

\author{J. E. Williams\( ^{1} \)}
\altaffiliation[Present address: ]{Electron and Optical Physics Division, National Institute of Standards and Technology, Geithersburg, MD 20899.}
\author{E. Zaremba\( ^{2} \)}
\author{B. Jackson\( ^{2} \)} 
\author{T. Nikuni\( ^{1} \)}
\author{A. Griffin\( ^{1} \)}

\address{\( ^{1} \)Department of Physics, University of Toronto, Toronto,
Ontario, Canada M5S 1A7}

\address{\( ^{2} \)Department of Physics, Queen's University, Kingston, Ontario,
Canada K7L 3N6}

\date{Sept. 10, 2001}

\begin{abstract}
\bigskip{}
We study surface modes of the condensate in the presence of a rotating
thermal cloud in an axisymmetric trap. By considering collisions that
transfer atoms between the condensate and noncondensate, we find that
modes which rotate in the same sense as the thermal cloud damp less
strongly than counter-rotating modes. We show that above a critical
angular rotation frequency, equivalent to the Landau stability criterion,
the co-rotating mode becomes dynamically unstable, leading to the
possibility of vortex nucleation. This kind of mechanism is proposed
as a natural explanation for the formation of vortices observed recently
in the experiment of Haljan \emph{et al} {[}P. C. Haljan \emph{et
al.}, cond-mat/0106362{]}. We also generalize our stability analysis
to treat the case where the stationary state of the condensate already
possesses a single vortex.
\end{abstract}
\maketitle
The recent success at producing vortices with quantized angular momentum
in trapped Bose gases \cite{Matthews1999,AboShaeer2001,Madison2000}
has raised interesting questions regarding the microscopic mechanisms
responsible for vortex nucleation. Several papers \cite{Dalfovo2000,Isoshima1999,Anglin,Muryshev,Feder2000,Tsubota}
have suggested the idea that a possible mechanism is the excitation
of low energy surface condensate modes, which have been studied in
various experiments. This can be understood \cite{Dalfovo2000,Anglin,Muryshev}
in terms of the well-known Landau criterion \cite{Landau1941}, which
predicts the excitation of condensate modes with frequency \( \omega _{m} \)
and angular momentum \( m\hbar  \) when the rotation rate \( \Omega  \)
of an external perturbation satisfies the condition \( \omega _{m}-m\Omega <0 \).
This defines a critical frequency, \( \Omega _{c,m}\equiv \omega _{m}/m \)
for transfer of angular momentum to the condensate \cite{Dalfovo2000,Isoshima1999},
where the particular mode excited depends upon the form of the ``stirring''
potential used. For example, a rotating trap anisotropy will predominantly
couple to a quadrupole surface mode, which will become dynamically
unstable above the Landau frequency \( \Omega _{c,2} \). In related
work, Sinha and Castin \cite{Sinha} have found dynamical instabilities
about stationary condensate solutions in the rotating frame \cite{Recati2001},
and link these to vortex nucleation in a recent experiment \cite{Madison2001}
. Further experiments lend support to the role of dynamical instabilities
of surface modes in the vortex formation process \cite{AboShaeer2001,Hodby}.

In the present letter, we extend this discussion of vortex nucleation
to finite temperatures and include for the first time the coupling
of the thermal cloud to the condensate. Our calculations address the
kind of experiments done at JILA \cite{Haljan}, in which a thermal
cloud is set into rotational motion above \( T_{\scriptstyle \rm BEC} \)
by means of an anisotropic rotating drive. A subsequent quench through
the Bose-Einstein condensation transition in the absence of the drive
then leads to the formation of a condensate containing vortices (when
the rotation frequency \( \Omega  \) of the thermal cloud exceeds
a critical value). What sets the JILA study \cite{Haljan} apart from
all other recent experiments on vortices \cite{Matthews1999,AboShaeer2001,Madison2001,Hodby}
is that the nucleation process occurs in the presence of an \textit{axially
symmetric trap}. The JILA experiment indicates that the rotating thermal
cloud is acting as a reservoir of angular momentum that is transferred
to the condensate in the process of vortex formation. It is clear
that a breaking of rotational symmetry is required if such a transfer
is to occur. Thermal excitation of condensate collective modes is
an obvious candidate for this symmetry breaking. The density fluctuations
associated with such a mode are coupled to the thermal cloud by means
of the condensate mean field, as well as through the collisional exchange
of atoms between the condensate and thermal cloud.

Of these two physical mechanisms, we focus on the latter in this paper.
The mean field coupling is also expected to be important, but we shall
demonstrate that the collisional exchange of atoms is already sufficient
to exhibit the dynamical instability of the condensate induced by
the rotating thermal cloud. Generalizing recent work by Williams and
Griffin \cite{Williams2001a}, we calculate the damping of condensate
collective modes at finite temperature due to the coupling via \( C_{12} \)
collisions between the condensate and noncondensate atoms. In formulating
this problem, we assume that the thermal cloud is in rigid body rotation
with angular frequency \( \Omega  \), as in the JILA experiments
\cite{Haljan}. Our major result is that the damping of the collective
modes of frequency \( \omega _{m} \) depends on the polarity \( m \)
of the collective mode with respect to the rotation of the thermal
cloud. As \( \Omega  \) is increased, the damping of the \( m<0 \)
modes increases while that of \( m>0 \) modes decreases. At a critical
angular frequency \( \Omega _{c,m}\equiv \omega _{m}/m \) of the
thermal cloud, the \( m>0 \) mode becomes unstable in the sense that
its amplitude grows exponentially with time. This signals a continuous
transfer of angular momentum from the thermal cloud to the condensate
which we identify with the onset of vortex formation.

The dynamics of the condensate wavefunction \( \Phi (\mathbf{r},t) \)
in the ZNG kinetic theory \cite{Zaremba1999} is described by a generalized
Gross-Pitaevskii (GP) equation, which at finite temperatures includes
the interaction with the noncondensate through a mean-field term and
a non-Hermitian damping term due to \( C_{12} \) collisions that
exchange atoms between the condensate and noncondensate. The noncondensate,
represented by a phase-space distribution function \( f(\mathbf{r},\mathbf{p},t) \)
obeying a semiclassical kinetic equation \cite{Zaremba1999}, is likewise
coupled to the condensate through mean field and collisional exchange
terms. The atoms are trapped in a static harmonic trap possessing
cylindrical symmetry \( U_{\mathrm{ext}}(\mathbf{r})=M\omega _{\perp }^{2}(\rho ^{2}+\lambda ^{2}z^{2})/2, \)
where the aspect ratio of the trap is \( \lambda =\omega _{z}/\omega _{\perp } \).
The noncondensate atoms move in a dynamic Hartree-Fock (HF) potential
\( U(\mathbf{r},t)=U_{\mathrm{ext}}(\mathbf{r})+2g[n_{c}(\mathbf{r},t)+\tilde{n}(\mathbf{r},t)] \),
where the interaction parameter \( g=4\pi \hbar ^{2}a/M \), \( a \)
is the \( s \)-wave scattering length, \( n_{c}(\mathbf{r},t)=|\Phi (\mathbf{r},t)|^{2} \)
and \( \tilde{n}(\mathbf{r},t) \) is the noncondensate density defined
by \( \tilde{n}({\textbf {r}},t)=\int d{\textbf {p}}f({\textbf {r}},{\textbf {p}},t)/(2\pi \hbar )^{3} \).

In our idealized model of the JILA experiments, the collective excitations
of the condensate are determined for an equilibrium state defined
by a thermal cloud undergoing solid body rotation at the angular frequency
\( \Omega  \). For this situation, the equilibrium distribution of
the rotating thermal cloud is given by\begin{equation}
\label{f0}
f^{0}(\mathbf{r},\mathbf{p})=\frac{1}{\exp \{\beta [(\mathbf{p}-M\mathbf{v}_{n0})^{2}/2M+U_{\mathrm{eff}}-\tilde{\mu }_{0}]\}-1},
\end{equation}
where the noncondensate velocity field is \( \mathbf{v}_{n0}(\mathbf{r})=\, \, \, \, \, \, \, \, \, \, \, \boldsymbol \Omega \times \mathbf{r}=\Omega \rho \boldsymbol {\hat{\phi }} \).
The effective potential acting on the thermal cloud is \( U_{\mathrm{eff}}(\mathbf{r})=U_{0}(\mathbf{r})-M\Omega ^{2}\rho ^{2}/2 \),
with \( U_{0}(\mathbf{r}) \) the equilibrium HF potential. The corresponding
equilibrium state of the condensate is given by the solution of the
time-independent generalized GP equation. Although we are primarily
interested in the initial onset of vortex formation, we can readily
extend our analysis to a situation in which the condensate already
has a single vortex centered on the \( z \)-axis with \( q \) units
of quantized circulation. The equilibrium condensate wavefunction
\( \Phi ^{(q)}(\rho ,\phi ,z)=\psi ^{(q)}(\rho ,z)e^{iq\phi } \)
is then defined by \begin{equation}
\left( -\frac{\hbar ^{2}\nabla ^{2}}{2M}+\frac{M}{2}v^{2}_{c0}+U_{\mathrm{ext}}+gn_{c0}+2g\tilde{n}_{0}\right) \psi ^{(q)}=\mu _{0}^{(q)}\psi ^{(q)}
\end{equation}
Here, the condensate velocity field is given by \( \mathbf{v}_{c0}(\rho )=(\hbar q/M\rho )\boldsymbol {\hat{\phi }} \).

In equilibrium, the chemical potentials of the thermal cloud and condensate
are related by \( \tilde{\mu }_{0}=\mu _{0}^{(q)}-\hbar \Omega q. \)
It is straightforward to prove that the form of \( f^{0}(\mathbf{r},\mathbf{p}) \)
in (\ref{f0}) is a stationary solution of the ZNG kinetic equation
for the case of a static axially symmetric trap when the chemical
potentials satisfy the above relation. Using (\ref{f0}), the thermal
cloud density is given by \( \tilde{n}_{0}(\rho ,z)=g_{3/2}[z_{0}(\rho ,z)]/\Lambda _{0}^{3}, \)
where the thermal de Broglie wavelength is \( \Lambda _{0}\equiv (2\pi \hbar ^{2}/Mk_{B}T)^{1/2} \)
and the equilibrium fugacity is \( z_{0}(\rho ,z)\equiv e^{-\beta (U_{\mathrm{eff}}(\rho ,z)-\tilde{\mu }_{0})} \).
As \( \Omega  \) increases, the aspect ratio of the noncondensate
density increases approximately as \( \omega _{z}/\sqrt{\omega _{\perp }^{2}-\Omega ^{2}} \)
.

Within our model, we neglect the dynamics of the noncondensate and
assume the thermal cloud remains rigidly rotating, as described by
(\ref{f0}) (see also Refs.~\cite{Williams2001a,Williams2001b,Duine}
for a discussion of this type of static approximation). In effect,
we are ignoring the perturbation of the rotating normal cloud induced
by the mean field of the oscillating condensate. Working with the
amplitude and phase of the condensate \( \Phi (\mathbf{r},t)=\sqrt{n_{c}(\mathbf{r},t)}e^{i\theta (\mathbf{r},t)} \),
the condensate dynamics is described by the coupled quantum hydrodynamic
equations \cite{Williams2001a,Zaremba1999}\begin{equation}
\label{dnc1}
\frac{\partial n_{c}}{\partial t}+\nabla \cdot (n_{c}\mathbf{v}_{c})=-\Gamma _{12}^{0}
\end{equation}
\begin{equation}
\label{dvc1}
M\frac{\partial \mathbf{v}_{c}}{\partial t}=-\nabla \left( \mu _{c}+\frac{1}{2}Mv_{c}^{2}\right) ,
\end{equation}
 where the local chemical potential \( \mu _{c}(\mathbf{r},t) \)
of the condensate is\begin{equation}
\label{muc}
\mu _{c}=-\frac{\hbar ^{2}}{2M}\frac{\nabla ^{2}n_{c}^{1/2}}{n_{c}^{1/2}}+U_{\mathrm{ext}}+gn_{c}+2g\tilde{n}_{0}.
\end{equation}
 The condensate velocity is \( \mathbf{v}_{c}(\mathbf{r},t)=(\hbar /M)\boldsymbol \nabla \theta (\mathbf{r},t) \).
The source term \( \Gamma _{12}^{0}(\mathbf{r},t) \) appearing in
(\ref{dnc1}) describes the collisional exchange of atoms between
the condensate and thermal cloud and is given by (see also Ref.~\cite{Williams2001a,Zaremba1999})
\begin{eqnarray}
\Gamma _{12}^{0} & \equiv  & \int \frac{d\mathbf{p}}{(2\pi \hbar )^{3}}C_{12}[f^{0},\Phi ]\nonumber \label{Gamma120} \\
 & = & \frac{n_{c}}{\tau _{12}}\left\{ e^{\beta [\varepsilon _{c}-\tilde{\mu }_{0}-M\mathbf{v}_{c}\cdot \mathbf{v}_{n0}]}-1\right\} ,\label{Gamma120} 
\end{eqnarray}
 where the collision time \( \tau _{12}(\mathbf{r},t) \) is defined
as\begin{equation}
\label{tau12}
\begin{array}{ccc}
\displaystyle \tau _{12}^{-1} & = & {\frac{2g^{2}}{(2\pi )^{5}\hbar ^{7}}}\int d\mathbf{p}_{1}\int d\mathbf{p}_{2}\int d\mathbf{p}_{3}\delta (\mathbf{p}_{c}+\mathbf{p}_{1}-\mathbf{p}_{2}-\mathbf{p}_{3})\\
 & \,  & \times \, \, \delta (\varepsilon _{c}+\tilde{\varepsilon }_{p_{1}}-\tilde{\varepsilon }_{p_{2}}-\tilde{\varepsilon }_{p_{3}})(1+f_{1}^{0})f_{2}^{0}f_{3}^{0}.\, \, \, \, \, \, \, \, \, \, \, \, \, \, \, \, \, \, \, \, 
\end{array}
\end{equation}
 The local condensate momentum per atom is \( \mathbf{p}_{c}(\mathbf{r},t)\equiv M\mathbf{v}_{c}(\mathbf{r},t) \)
and the local energy is \( \varepsilon _{c}(\mathbf{r},t)=\mu _{c}(\mathbf{r},t)+M\mathbf{v}_{c}^{2}/2. \)
The equilibrium single-particle distribution function \( f_{i}^{0}\equiv f^{0}(\mathbf{r},\mathbf{p}_{i}) \)
is as given by (\ref{f0}). We note that the term \( \Gamma _{12}^{0} \)
in (\ref{dnc1}) is the only source of damping in our calculations.
Our static approximation for the rotating thermal cloud precludes
Landau damping, which can be expected to provide additional damping
of the condensate modes.

To study the collective modes, we consider the density \( \delta n_{c}(\mathbf{r},t) \)
and velocity \( \delta \mathbf{v}_{c}(\mathbf{r},t) \) fluctuations
of the condensate about equilibrium \( n_{c}(\mathbf{r},t)=n_{c0}(\mathbf{r})+\delta n_{c}(\mathbf{r},t) \)
and \( \mathbf{v}_{c}(\mathbf{r},t)=\mathbf{v}_{c0}(\mathbf{r})+\delta \mathbf{v}_{c}(\mathbf{r},t) \).
In the Thomas-Fermi (TF) limit, the linearized equations of motion
for the condensate fluctuations are given by \begin{equation}
\label{dnc2}
\frac{\partial \delta n_{c}}{\partial t}+\nabla \cdot \left( n_{c0}\delta \mathbf{v}_{c}+\mathbf{v}_{c0}\delta n_{c}\right) =-\delta \Gamma _{12}
\end{equation}
\begin{equation}
\label{dvc2}
\frac{\partial \delta \mathbf{v}_{c}}{\partial t}=-\nabla \left( \frac{g}{M}\delta n_{c}+\mathbf{v}_{c0}\cdot \delta \mathbf{v}_{c}\right) .
\end{equation}
 It is straightforward to show that the linearized form of \( \Gamma _{12}^{0}(\mathbf{r},t) \)
given in (\ref{Gamma120}) is \begin{equation}
\label{dGam1}
\delta \Gamma _{12}=\frac{1}{\tau '}\left[ \delta n_{c}+\frac{M}{g}\delta \mathbf{v}_{c}\cdot (\mathbf{v}_{c0}-\mathbf{v}_{n0})\right] ,
\end{equation}
 where \( 1/\tau '(\mathbf{r})\equiv gn_{c0}(\mathbf{r})/[k_{B}T\tau _{12}^{0}(\mathbf{r})]. \)
Here \( \tau _{12}^{0}(\mathbf{r}) \) is given by the expression
in (\ref{tau12}), but with the dynamic or time-dependent condensate
quantities replaced by the corresponding equilibrium ones. We remark
that the second term in (\ref{dGam1}) has the form of a mutual friction
term depending on the relative velocity \( \mathbf{v}_{c0}(\mathbf{r})-\mathbf{v}_{n0}(\mathbf{r}) \)
between the condensate and normal gas, which couples to the velocity
fluctuation of the condensate \( \delta \mathbf{v}_{c} \). For \( \Omega =0 \)
and \( \mathbf{v}_{c0}=0 \), the results in (\ref{dnc1})-(\ref{tau12})
reduce to those in Ref.~\cite{Williams2001a}.

In the TF approximation, the condensate density is given by \( n_{c0}(\mathbf{r})=(\mu _{0}^{(q)}-U_{\mathrm{ext}}-2g\tilde{n}_{0}-\frac{1}{2}Mv_{c0}^{2})/g. \)
To simplify the calculation, we neglect the effect of the noncondensate
mean field on the condensate profile \cite{Williams2001a}. In addition,
we neglect the effect of the vortex core on the overall shape of the
condensate. This approximation is valid when \( \xi ^{2}/R_{0}^{2}\ll 1 \),
where \( \xi  \) is the healing length and \( R_{0} \) is the Thomas-Fermi
condensate radius \cite{Svidzinsky1998}. We therefore approximate
the condensate density as \( n_{c0}=(\mu _{TF}-U_{\mathrm{ext}})/g \),
where \( \mu _{TF}(T)=\hbar \omega _{\perp }[15\lambda N_{c}(T)a/d_{\perp }]^{2/5}/2 \)
and \( d_{\perp }=\sqrt{\hbar /M\omega _{\perp }} \) . 

In an axially symmetric trap, the density and velocity fluctuations
of the collective modes take the form \( \delta n_{c}(\mathbf{r},t)=\delta n_{m}(\rho ,z)e^{i(m\phi -\omega _{m}t)} \)
and \( \delta \mathbf{v}_{c}(\mathbf{r},t)=(\hbar /M)\nabla \delta \theta _{m}(\rho ,z)e^{i(m\phi -\omega _{m}t)} \).
Substituting these expressions into (\ref{dnc2}) and (\ref{dvc2})
gives \begin{eqnarray}
\hbar \left[ \hat{S}+\frac{im}{\tau '}\left( \frac{V}{\rho }-\Omega \right) \right] \delta \theta _{m} &  & \nonumber \\
-ig\left( \omega _{m}-\frac{mV}{\rho }+\frac{i}{\tau '}\right) \delta n_{m} & = & 0\label{modes1} 
\end{eqnarray}
\begin{equation}
\label{modes2}
-i\hbar \left( \omega _{m}-\frac{mV}{\rho }\right) \delta \theta _{m}+g\delta n_{m}=0,
\end{equation}
 where \( \hat{S}\equiv g[\nabla \cdot (n_{c0}\nabla )-m^{2}n_{c0}/\rho ^{2}]/M \)
and \( V\equiv \hbar q/M\rho  \). These equations can be combined
into a single equation for \( \delta n_{m} \), following the procedure
used in Ref.~\cite{Svidzinsky1998}. We relate \( \delta \theta _{m} \)
to \( \delta n_{m} \), making use of the following approximation
to (\ref{modes2}) \begin{equation}
\label{densphaserel}
\delta \theta _{m}\simeq -i\frac{g}{\hbar \omega _{m}}\left( 1+\frac{mV}{\omega _{m}\rho }\right) \delta n_{m}.
\end{equation}
 This is valid for \( \rho \gg \xi  \), consistent with neglecting
the effect of the vortex core on the static condensate density profile.
We then obtain a single wave equation for the condensate density fluctuation
for mode \( m \)\begin{equation}
\label{Eigeq1}
(\omega _{m}^{2}+\hat{S}+\hat{S}_{q}+i\hat{S}_{\tau })\delta n_{m}=0,
\end{equation}
 where \begin{equation}
\label{Sq}
\hat{S}_{q}\equiv \frac{2m\hbar q}{\omega _{m}M\rho ^{2}}\left[ \omega _{\perp }^{2}-\omega _{m}^{2}+\frac{2gn_{c0}}{M\rho }\left( \frac{1}{\rho }-\frac{\partial }{\partial \rho }\right) \right] 
\end{equation}
and \begin{equation}
\label{Sgamma}
\hat{S}_{\tau }\equiv (\omega _{m}-m\Omega )\frac{1}{\tau '}.
\end{equation}

In order to obtain an approximate solution of (\ref{Eigeq1}), we
treat the effect of the vortex \( \hat{S}_{q} \) and the damping
\( \hat{S}_{\tau } \) perturbatively \cite{Williams2001a} by expanding
in the zeroth-order TF solutions given by \( \hat{S}\delta n_{S}=-\omega _{S}^{2}\delta n_{S} \)
\cite{Stringari1996}. Eq. (\ref{Eigeq1}) can then be simplified
to \cite{Williams2001a} \begin{equation}
\label{perurbEigeq}
\omega _{m}^{2}-[\omega _{m}^{(q)}]^{2}+i(\omega _{m}-m\Omega )\left\langle \frac{1}{\tau '}\right\rangle =0,
\end{equation}
 where the spatial average of an operator \( \chi  \) is defined
by \( \langle \chi \rangle \equiv \int d\mathbf{r}\chi (\mathbf{r})\delta n_{S}^{2}(\mathbf{r})/\int d\mathbf{r}\delta n_{S}^{2}(\mathbf{r}). \)
The frequency \( \omega _{m}^{(q)} \) is the collective mode frequency
of a vortex state \cite{Svidzinsky1998,Zambelli1998}\begin{equation}
\label{omegaq}
[\omega _{m}^{(q)}]^{2}\equiv \omega _{S}^{2}(1+qm\Delta _{m}),
\end{equation}
 where \( \omega _{S}\equiv \omega _{m}^{(0)} \) is the TF frequency
for \( q=0 \), with \begin{equation}
\label{Delta}
\Delta _{m}\equiv \frac{2\hbar }{\omega _{S}^{3}M}\left\langle (\omega _{S}^{2}-\omega _{\perp }^{2})\frac{1}{\rho ^{2}}+\frac{2g}{M}\frac{n_{c0}}{\rho ^{3}}\left( \frac{\partial }{\partial \rho }-\frac{1}{\rho }\right) \right\rangle .
\end{equation}
 The solution of (\ref{perurbEigeq}) to lowest order in the damping
is \( \omega _{m}=\omega _{m}^{(q)}-i\Gamma _{m}^{(q)} \), where
the damping rate is given by \begin{equation}
\label{Imag}
\Gamma _{m}^{(q)}=\frac{1}{2}\left\langle \frac{1}{\tau '}\right\rangle \left( 1-\frac{m\Omega }{\omega _{m}^{(q)}}\right) .
\end{equation}

We note that the finite \( T \) result in (\ref{omegaq}) is formally
identical to the \( T=0 \) result in Ref.~\cite{Svidzinsky1998}.
The only difference is the use of \( N_{c}(T) \) for the number of
condensate atoms. As discussed in Refs.~\cite{Svidzinsky1998,Zambelli1998},
collective modes with opposite polarity \( m \) are split in frequency
since the vortex circulation breaks the azimuthal symmetry. Modes
rotating in the same sense as the vortex (\( m>0 \)) are shifted
higher in frequency and modes rotating against the flow of the vortex
(\( m<0 \)) are shifted down, in agreement with experiments \cite{Madison2000,Haljan}.

The expression for the damping (\ref{Imag}) of the collective modes
of a condensate interacting with a rigidly rotating thermal gas is
our main new result. Interestingly, it predicts that modes of opposite
polarity \( m \) are affected by the rotating thermal cloud in quite
distinct ways. The damping of modes that rotate against the flow of
the thermal cloud \( (m<0) \) increases linearly with angular frequency
\( \Omega  \), while the damping of modes that rotate in the same
sense as the thermal cloud \( (m>0) \) decreases with increasing
\( \Omega  \). This prediction should be easily observable. For \( m>0 \)
excitations, the damping in (\ref{Imag}) goes to zero when \( \Omega  \)
reaches a critical value defined by \begin{equation}
\label{Omegac}
\Omega _{c,m}^{(q)}\equiv \frac{\omega _{m}^{(q)}}{m}.
\end{equation}
 For \( \Omega >\Omega _{c,m}^{(q)} \), the damping changes sign,
indicating the onset of a dynamical instability (i.e., \( \delta n_{m}\sim e^{+|\Gamma _{m}|t} \)).

The result in (\ref{Omegac}) is equivalent to the usual Landau criterion
for the excitation of surface modes by an external anisotropic perturbation
as discussed in Ref.~\cite{Dalfovo2000}. In contrast, our finite-temperature
mechanism is effective even in an axisymmetric trap. Physically, one
would expect all of the surface modes to be thermally occupied, so
that for a noncondensate rotation rate of \( \Omega  \), instabilities
can be induced in any of these modes so long as \( \Omega >\Omega _{c,m}^{(q)} \).
In other words, at least one mode will be unstable when \( \Omega  \)
exceeds a critical frequency defined by the minimum of a plot of \( \omega _{m}^{(q)}/m \)
versus \( m \) \cite{Dalfovo2000}. This critical rotation frequency
is generally larger than that required to ensure thermodynamic stability
of a vortex state \cite{Isoshima1999}, so that it is likely that
the primary role of surface mode instabilities is to facilitate tunneling
of one or more vortex lines into the condensate bulk. This has recently
\cite{Kramer} been demonstrated using energy arguments for the particular
case of a \( m=2 \) quadrupole mode. When a single \( q=1 \) vortex
is already present, our result (18) shows that the critical rotation
frequencies increase due to corresponding upward shifts in the \( m>0 \)
frequencies \cite{Svidzinsky1998,Zambelli1998}. Again, taking the
\( m=2 \) mode as a concrete example, our TF analysis predicts \( \Omega _{c,2}^{(0)}=\omega _{\bot }/\sqrt{2} \)
without a vortex, while \( \Omega _{c,2}^{(1)}=\omega _{\bot }\sqrt{(1+2\Delta _{2})/2} \)
for a \( q=1 \) vortex state. After nucleation, the vortices will
eventually equilibrate into a thermodynamically stable lattice configuration,
as discussed in \cite{Feder1999b,Tsubota,Feder2001b}.

We emphasize that the present paper only evaluates the damping from
\( C_{12} \) collisions in the {}``collisionless'' or mean field
region. We expect that Landau damping, which would arise if we included
thermal cloud fluctuations \cite{Williams2001a,Jackson}, would also
exhibit the same dynamical instability for \( \Omega >\Omega _{c,m}^{(q)} \).
The damping term in (\ref{dGam1}) can be interpreted as a kind of
mutual friction, although it is quite different from the usual mutual
friction arising in the two-fluid hydrodynamic domain, such as discussed
in Ref.~\cite{Fedichev}.

In summary, stimulated by recent work at \( T=0 \) on the nucleation
of vortices by a rotating anisotropic harmonic potential, we have
generalized the discussion to finite temperatures and considered the
case where the driving field is a rigidly rotating thermal cloud \cite{Haljan}.
Our calculation is explicitly based on the collisional exchange coupling
between this rotating thermal cloud and the condensate which gives
rise to mode damping. This damping can change sign when the thermal
cloud angular frequency \( \Omega  \) reaches a critical value given
by the Landau criterion for excitation of collective modes of angular
momentum \( m \) along the \( z \) axis. Importantly, our result
also applies to the stability of a singly quantized vortex against
excitations of surface collective modes, where it is associated with
dynamic nucleation of additional vortices.

We thank the authors of Ref.~\cite{Haljan} for stimulating discussions
about the JILA experiment and David Feder for a critical reading of
an early draft of this paper. This research was supported by NSERC
of Canada and JSPS of Japan.

\end{document}